\documentclass[aps,pra,twocolumn,showpacs,preprintnumbers,amsmath,amssymb,floatfix]{revtex4}

\usepackage{graphicx}
\usepackage{graphics}

\usepackage{bm}
\begin{document}

\title{Tunable anisotropic magnetism in trapped two-component Bose gases}
\author{Yongqiang Li$^{1,2}$, M. Reza Bakhtiari$^{1}$, Liang He$^{1}$, Walter Hofstetter$^{1}$}
\affiliation{$^{1}$Institut f\"ur Theoretische Physik, Johann Wolfgang
Goethe-Universit\"at, 60438 Frankfurt/Main, Germany\\
$^{2}$Department of Physics, National University of Defense Technology, Changsha 410073, P. R. China}
\date{\today}

\begin{abstract}
We theoretically address magnetic ordering at zero and finite
temperature in both homogeneous and trapped Bose-Bose mixtures in
optical lattices. By using Bosonic Dynamical Mean-Field Theory, we
obtain the phase diagram of the homogeneous two-component
Bose-Hubbard model in a three-dimensional (3D) cubic lattice, which
features competing magnetic order of XY-ferromagnetic and
anti-ferromagnetic type in addition to the Mott and superfluid
states. We show that these magnetic phases persist also in the
presence of a harmonic trap.

\end{abstract}


\maketitle
\section{Introduction}
Quantum magnetism is one of the most intriguing areas in condensed-matter physics. Even though the attempts to understand it root back to the early days of quantum theory, still there are many open questions even at the level of the minimalistic Hubbard model \cite{Auerbach}. Many theoretical and experimental efforts have been devoted to revealing the mechanisms behind magnetic ordering of many-body systems \cite{Sachdev2008}. Due to the high level of complexity in solid-state systems, a quantitative comparison between theory and experiment seems a very challenging task, if not impossible at all. Therefore it is highly desirable to work with systems which are able to \textit{simulate} the original solid-state many-body systems, but in a much more controllable way.

Over the past decade cold-atomic quantum gases in optical lattices
have provided an excellent laboratory for investigating many-body
quantum systems with an unprecedented level of precision and
control. Nowadays one can routinely create optical lattices with
different geometries that mimic solid-state crystalline structures
and then load different quantum gases into them, playing the role of
electrons. Based on this experimental progress, fundamental
many-body phenomena such as the Mott insulator quantum phase
transition of interacting bosons have been realized
\cite{Greiner2002}. For bosonic gases in optical lattices correlated
atom tunneling \cite{Folling2007} and superexchange due to
second-order atom tunneling \cite{Trotzky2008} have been observed,
which are the basic mechanisms leading to quantum magnetism. At the
current stage, detecting anti-ferromagnetic long-range ordering of
spinful fermions or bosons is arguably the most challenging goal and
there are large experimental and theoretical efforts directed
towards reaching this phase. This may eventually give insights into
the mechanism of high-$T_c$ superconductivity as well
\cite{Esslinger2010}. The recently developed optical quantum gas
microscope \cite{Bakr2009,Sherson2010} offers the possibility to
detect quantum magnetism on the single-atom level, and has been
utilized to observe the phase transition of a one-dimensional chain
of interacting Ising spins by using a Mott insulator of spinless
bosons in a tilted optical lattice \cite{Simon_2011}.

In parallel to these experimental developments, there are several
theoretical proposals how to gain fundamental insight into quantum
magnetism via ultracold gases
\cite{Kuklov2003,Duan2003,Altman2003,Capogrosso2009,Powell2009,Capogrosso2010,Parny2010,Barthel2009,Noda2009,Shrestha2010}.
One of the aims has been to realize a spin Hamiltonian by
interacting ultracold bosons \cite{Kuklov2003, Duan2003}. For this
two-component Bose gas in an optical lattice, which is effectively
described by the Bose-Hubbard model, the complete phase diagram was
mapped out by a variational approach \cite{Altman2003}.

A related important experimental achievement is given by mixtures of
different quantum gases, whether with different statistics
(Bose-Fermi mixture) or different isotopes of the same type
(Bose-Bose or Fermi-Fermi mixture). Recently a Bose-Bose mixture of
$^{87}$Rb and $^{41}$K in a 3D optical lattice has been realized
\cite{Catani2008}.  Here it is possible to tune independently the
inter-species and intra-species interactions. In a follow-up
experiment the same Bose-Bose mixture has been used to investigate
the entropy exchange between two spin-dependent traps though without
lattice \cite{Catani2009}. Another Bose-Bose mixture, made of two
hyperfine states of $^{87}$Rb, has served as a spin gradient
thermometer which allows to measure the temperature of ultra-cold
atoms in optical lattices \cite{Weld2009,Medley2010}. In a further
experiment the effect of the Bose-Bose inter-species interaction on
the bosonic superfluidity of one of the components in a 3D optical
lattice has been explored \cite{Gadway2010}.

Motivated by these experiments, here we theoretically investigate a
two-component Bose gas in 2D and 3D optical lattices. This system
can be effectively modeled by the Bose-Hubbard model. We will
specifically consider the case of filling $n=1$ and $n=2$ per site.
We investigate the homogeneous (untrapped) system by means of
Bosonic Dynamical Mean-Field Theory (BDMFT)
\cite{Byczuk2008,Hubener2009,Hu2009,Snoek2010,Werner2010} and the
harmonically trapped case by its real-space generalization (RBDMFT),
which extends the original BDMFT formalism to the study of
inhomogeneous systems. DMFT has been established as a powerful tool
for strongly-correlated fermionic systems \cite{Georges1996}, while
BDMFT has recently been shown to provide a qualitatively and in 3D
even quantitatively accurate picture of the Bose-Hubbard model
\cite{Werner2010}. For the homogeneous system, we extend our
previous BDMFT analysis to higher filling $n=2$ and a different set
of interaction parameters motivated by experiments. We map out the
phase diagram and obtain diverse phases such as superfluid,
unordered Mott state and XY-ferromagnetic order. In addition we turn
to the inhomogeneous (trapped) Bose-Hubbard model which is more
closely related to the experimental situation. We include the effect
of the external confining potential by RBDMFT, which  assumes
site-dependent self-energies. In parallel we perform also a
complementary calculation based on a Local Density Approximation
(LDA) combined with BDMFT which is computationally more affordable.
Comparing results of both methods, we examine the magnetic
properties of the system for a wide range of parameters. To our best
knowledge this is the first systematic and non-perturbative study of
the magnetic properties of a two-component inhomogeneous
Bose-Hubbard model. It will bring more insight into ongoing
experiments on Bose-Bose mixtures in optical lattices.

The paper is organized as follows: in section II we give a detailed
description of the RBDMFT approach. Section III covers our results
for the homogeneous Bose-Hubbard model. In Section IV we consider
the trapped system and present results by RBDMFT and LDA+BDMFT. We
summarize with a discussion in Section V.

\section{Model and Method} \label{sec:model}
We consider a two-component bosonic mixture in an optical lattice
with either 2D square or 3D cubic geometry. Experimentally the
Bose-Bose mixture could consist of two different species,
\textit{e.g.} $^{87}$Rb and $^{41}$K as in Ref. \cite{Catani2008} or
two different hyperfine states of a single species,  \textit{e.g.}
$^{87}$Rb as in Refs. \cite{Trotzky2008,Weld2009,Gadway2010}. In
addition we include an external harmonic trapping potential which
gives rise to inhomogeneity. This system can be described by a
two-component inhomogeneous Bose-Hubbard model
\begin{eqnarray} \label{Hamil}
 \mathcal{H}=&-& \sum_{\stackrel{<i,j>}{\nu=b,d}} t_\nu (b^\dagger_{i\nu}b_{j\nu}+h.c)+\frac{1}{2}\sum_{i,\lambda\nu} U_{\lambda\nu} \hat{n}_{i,\lambda}
(\hat{n}_{i,\nu}-\delta_{\lambda\nu}\nonumber) \\
&+&\sum_{i,\nu=b,d} (V_i-\mu_\nu)\hat{n}_{i\nu}
\end{eqnarray}
In this Hamiltonian $\langle i,j\rangle$ represent the nearest
neighbor sites $i,j$ and we denote the two bosonic species as $b,d$
which are labeled by the index $\lambda (\nu)=b, d$. The bosonic
creation (annihilation) operator for species $\nu$ at site $i$ is
$b^{\dagger}_{i\nu}$ ($b_{i\nu}$) and the local density is
$\hat{n}_{i,\nu}=b^\dagger_{i\nu}b_{i\nu}$. Due to possibly
different masses or a spin-dependent optical lattice, these two
species in general hop with non-equal amplitudes $t_b$ and $t_d$.
$U_{\lambda\nu}$ denotes the inter- and intra-species interactions,
which can be tuned via a Feshbach resonance or by spin-dependent
lattices. $\mu_\nu$ denotes the global chemical potential for the
two bosonic species and $V_i$ the harmonic trap.

For the fermionic Hubbard model, DMFT has been developed and
implemented successfully \cite{first_paper, Georges1996} as a
non-perturbative formalism to study strongly-correlated electronic
systems. This includes real-space generalizations of DMFT to address
inhomogeneous fermionic systems \cite{Snoek2008,Helmes2008}. In
spite of this success, a bosonic version of DMFT (BDMFT) for the
Bose-Hubbard model has been formulated \cite{Byczuk2008} and
implemented \cite{Hubener2009,Hu2009, Werner2010} only very
recently. Inspired by the case of fermions, here we extend BDMFT
\cite{Hubener2009} to a real-space BDMFT (RBDMFT) formalism for
inhomogeneous systems to include also the harmonic trap which is
crucial in the experiments. RBDMFT is capable of providing an
accurate and non-perturbative description of the ground-state of the
inhomogeneous Bose-Hubbard model (\ref{Hamil}). Like fermionic DMFT,
also BDMFT assumes that the self-energy of the system is local.
However, in an inhomogeneous system it depends on the lattice site,
\textit{i.e.}
$\mathbf{\Sigma}^{(i,j)}_{\lambda\nu}=\mathbf{\Sigma}^{(i)}_{\lambda\nu}\,\delta_{i,j}$
where $\delta_{i,j}$ is a Kronecker delta.

In a more formal language, first we map the Hamiltonian (\ref{Hamil}) onto a set of individual single-site problems each of which is described by a
\textit{local} effective action
\begin{eqnarray}\label{action}
 S_{\text{eff}}^{(i)}&=&\!\!\!\int_0^\beta\!\!\! d\tau\,d\tau'\!\!\!\sum_{\lambda,\nu=\{b,d\}}\mathbf{b}_{\lambda}^{(i)}(\tau)^\dagger\boldsymbol{\mathcal{G}}_{0,\lambda\nu}^{(i)}(\tau-\tau')^{-1}
\mathbf{b}_{\nu}^{(i)}(\tau')\nonumber \\
&+&\int_0^\beta d\tau\Big\{\sum_{\lambda,\nu} \frac{1}{2} U_{\lambda,\nu}n^{(i)}_{\lambda}(\tau)\Big(n^{(i)}_{\nu}(\tau)-\delta_{\lambda\nu}\Big)\label{eff_action} \nonumber \\
&-&\sum_{\langle ij\rangle , \nu} t_\nu \Big(b_{\nu}^{(i)}(\tau)^{*}\phi^{(i)}_{j,\nu}(\tau)+b_{\nu}^{(i)}(\tau)\phi^{(i)}_{j,\nu}(\tau)^{*}\Big)\Big\}
\end{eqnarray}
where $i$ is the index of the lattice site. In this equation $\tau$ is imaginary time and the function $\boldsymbol{\mathcal{G}}_{0,\lambda\nu}^{(i)}(\tau-\tau')$ is a local non-interacting propagator interpreted as a \textit{local} dynamical Weiss mean-field  and is determined in a self-consistent manner. Here we use the Nambu notation $\mathbf{b}_{\nu}^{(i)}(\tau)\equiv \big(b_{\nu}^{(i)}(\tau),b_{\nu}^{(i)}(\tau)^*\big)$. Moreover the static bosonic mean-fields are defined in terms of the bosonic operator
$b_{j,\nu}$ as
\begin{equation} \label{phi_def}
\phi^{(i)}_{j,\nu}(\tau)=\langle b_{j,\nu}\rangle_0.
\end{equation}
The index 0 indicates that all averages are taken for the cavity system \textit{i.e.} excluding the impurity site. Now each of
the local actions can be treated as an impurity in the presence of a bath (representing the influence of the rest of the lattice) and therefore captured via an Anderson impurity Hamiltonian. There are several techniques to solve the impurity model. Here we apply Exact Diagonalization (ED) \cite{Hubener2009}.

In practice, we start with an initial set of local Weiss Green's functions and local bosonic superfluid order parameters $\phi^{(i)}_{j,\nu}(\tau)$. After solving the action (\ref{action}), we obtain a set of local self-energies $\mathbf{\Sigma}^{(i)}_{\lambda\nu}(i\omega_n)$ with $\omega_n$ being Matsubara frequency. Then we employ the Dyson equation in real-space representation in order to compute the interacting lattice Green's function
\begin{equation} \label{G}
 \mathbf{G}(i\omega_n)^{-1}=\mathbf{G}_0(i\omega_n)^{-1}-\mathbf{\Sigma}(i\omega_n).
\end{equation}
The site-dependence of the Green's functions is shown by boldface quantities that denote a matrix form with site-indexed elements.
Here $\mathbf{G}_0(i\omega_n)^{-1}$ stands for the non-interacting Green's function
\begin{equation}\label{G0}
 \mathbf{G}_0(i\omega_n)^{-1}=(\mu+i\omega_n)\mathbf{1}-\mathbf{t}-\mathbf{V}.
\end{equation}
In this expression, the matrix elements $t_{ij}$ are hopping amplitudes for a given lattice structure and the external potential is included via $V_{ij}=\delta_{ij}V_i$ with $V_i=V_0r_i^2$ and $r_i$ being the distance from the trap center. Eventually the self-consistency loop is closed by specifying the Weiss Green's function via the local Dyson equation
\begin{equation}\label{g0}
 \Big(\boldsymbol{\mathcal{G}}^{(i)}_{0,\lambda\nu}(i\omega_n)\Big)^{-1}=\Big( \mathbf{G}^{(i)}_{\lambda\nu}(i\omega_n)\Big)^{-1} + \mathbf{\Sigma}^{(i)}_{\lambda\nu}(i\omega_n).
\end{equation}
where the diagonal elements of the lattice Green's function yield the interacting local Green's function
$
\mathbf{G}^{(i)}_{\lambda\nu}(i\omega_n)= (\mathbf{G}_{\lambda,\nu}(i\omega_n))_{ii}
$.
This self-consistency loop is repeated until the desired accuracy for superfluid order parameters and Weiss Green function is obtained.

Complementary to RBDMFT, in this work we employ an LDA approach
combined with single-site BDMFT to explore the physics of the model
(\ref{Hamil}). Here we adjust the chemical potential on each lattice
site according to LDA as $\mu_\nu(r)=\mu_\nu^0-V(r)$. The advantage
of this approach is the larger system size accessible. We validate
it by comparison with the more rigorous RBDMFT approach. In this
paper, we apply RBDMFT for 2D lattices and LDA+BDMFT both for 2D and
3D lattices.

\section{Magnetic phases of a Homogeneous Bose-Hubbard model}
We start by exploring the two-component Bose-Hubbard model in the
homogeneous case. We consider a 3D cubic lattice and focus  on the
situation of total filling $n=n_b+n_d$ being $n=1$ and $n=2$ with
balanced densities $n_b=n_d=0.5$ and $n_b=n_d=1$ respectively. For
each filling, we calculate both zero and finite temperature phase
diagrams. We focus on the interaction regime where the inter-species
interactions, $U_{bb}\equiv U_b, U_{dd}\equiv U_d$, and
intra-species interaction, $U_{bd}$, are just slightly different,
\textit{i.e.} $U_{b}=U_{d}=1.01\, U_{bd}$. This particular regime of
interactions is accessible by Feshbach resonances, and indeed our
choice is motivated by a recent experiment at MIT \cite{Weld2009},
where a sample of $^{87}$Rb atoms in hyperfine states $|1, -1
\rangle$ and $|2, -2 \rangle$ with nearly equal inter- and
intra-species interactions has been prepared. The selection of
interactions ($U_{bd}$ slightly smaller than $U_{b,d}$) is due to
more novel magnetic phases appearing in this regime
\cite{Altman2003}. In all our calculations we set $U_{bd}=1$ as the
unit of energy, and $z$ is the number of the nearest neighbors for
each site.

\begin{figure}
\begin{center}
\includegraphics[clip,width=.9\linewidth]{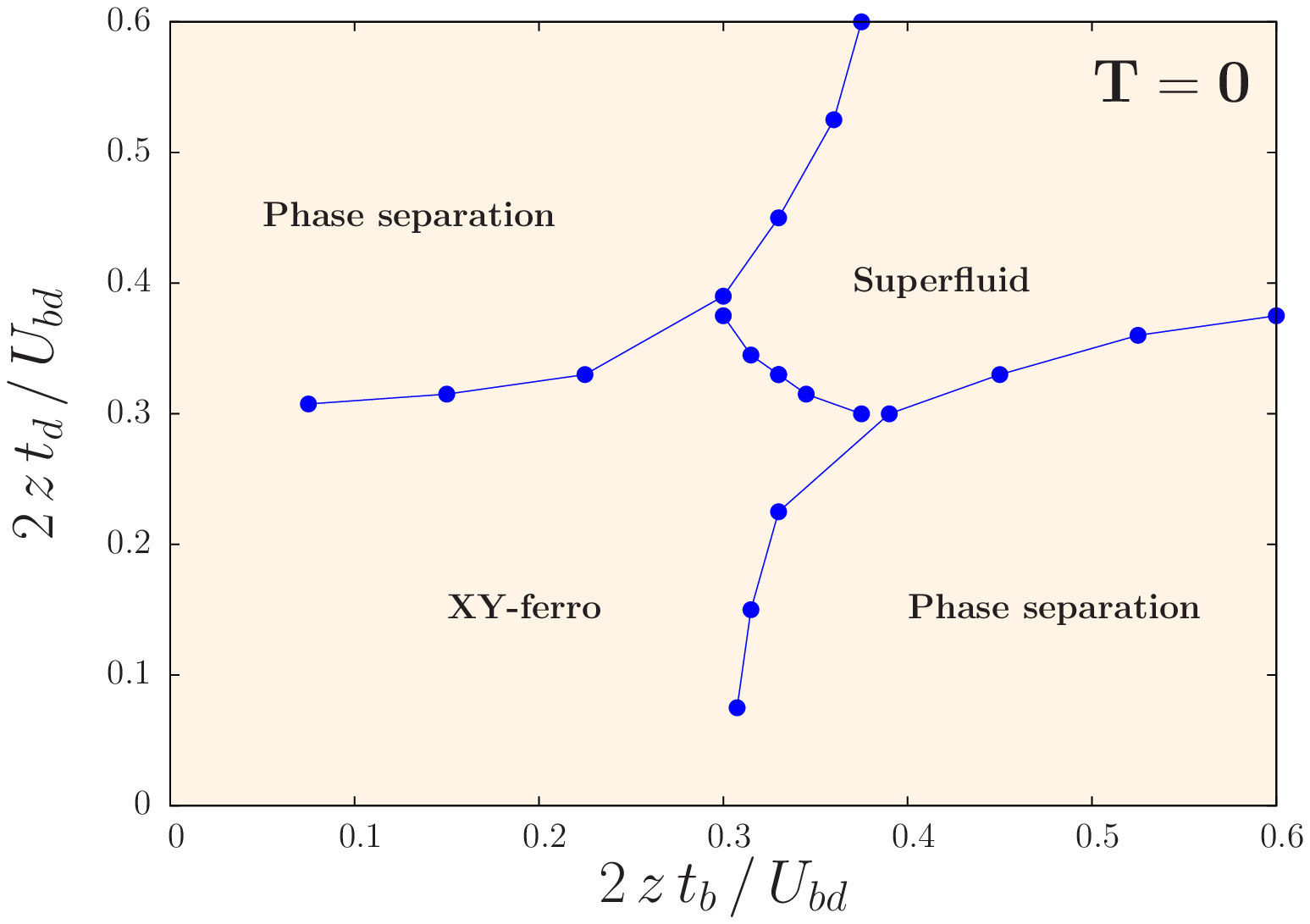}
\includegraphics[clip,width=.9 \linewidth]{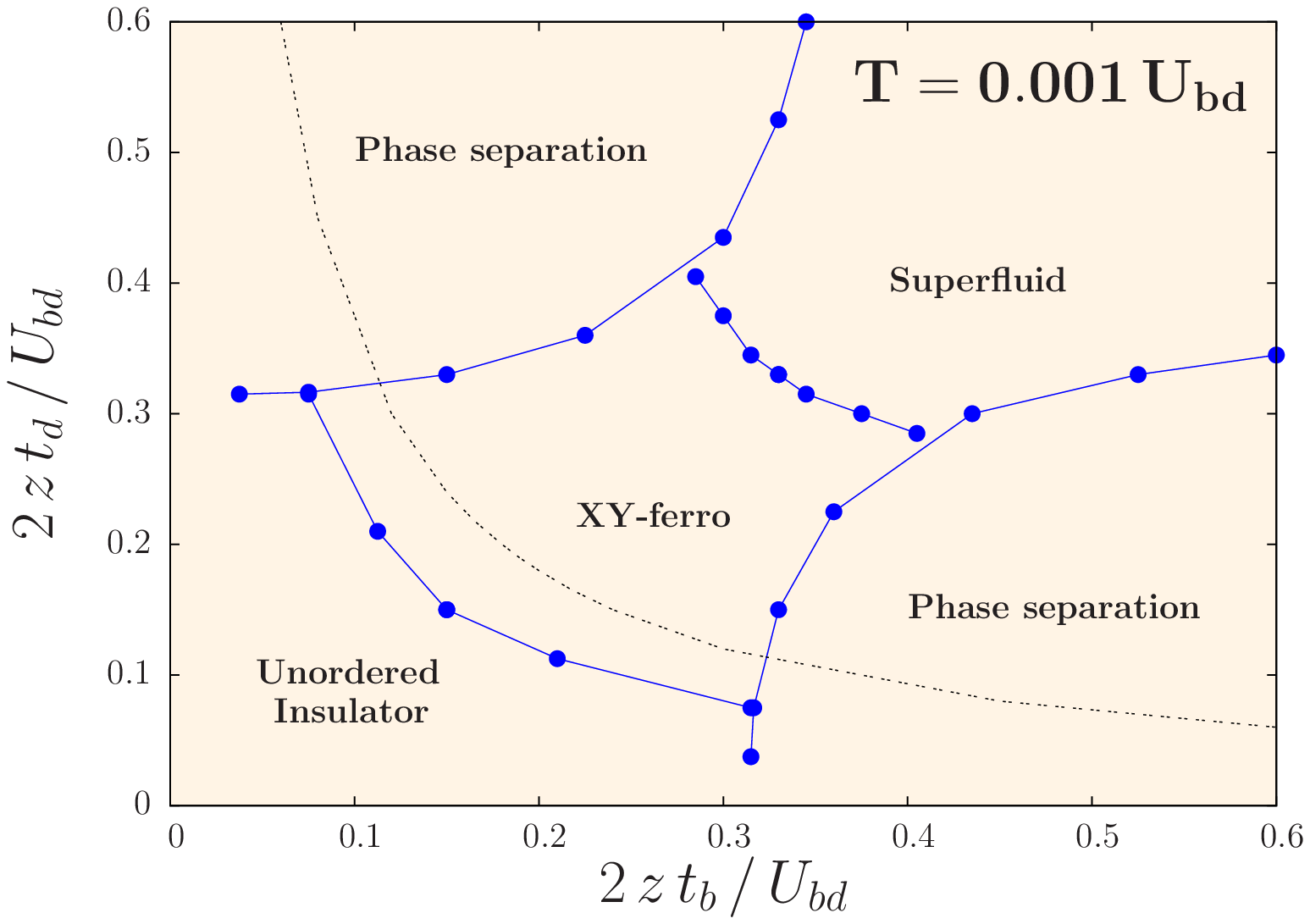}
\caption{ Upper panel: zero-temperature phase diagram for the
two-component Bose-Hubbard model in a  3D cubic lattice, as a
function of hopping parameters. The interaction values are  $U_{b}=
U_{d}=1.01\,U_{bd}$ and the total filling is $n=1$ with
$n_b=n_d=0.5$ (except in the phase separation regime).
 Lower panel: finite temperature phase diagram ($T=0.001 U_{bd}$). The energy scale of the magnetic coupling $4t_bt_d/U_{bd}$
is shown by the black dashed line.} \label{fig1new}
\end{center}
\end{figure}

Fig.~\ref{fig1new} displays the zero and finite temperature phase
diagrams of the system with a total filling of one particle per
site, $n=1$. At $T=0$ (upper panel) we find three distinct phases
which are characterized  according to the value of the superfluid
order parameters $\phi_b , \phi_d$ and the two-body correlator
$\phi_{bd}=\langle b\,d^\dagger\rangle- \langle b \rangle \langle
d^\dagger \rangle
>0 $ which indicates the XY-ferromagnetic
spin-ordering. When both species have comparably large hopping, we
find a superfluid phase characterized by $\phi_{b,d}\neq0$. Instead
when the hopping amplitudes are very different, the species with
larger hopping is more easily delocalized and therefore superfluid,
while the other component favors a Mott insulating phase. In this
parameter regime, we do not find a homogeneous converged BDMFT
solution where each component has the same filling, which indicates
phase separation between the superfluid and the Mott insulator. We
notice that the phase diagram is symmetric upon particle interchange
and this symmetry is also manifested in the Hamiltonian
(\ref{Hamil}). The third phase emerges when the hopping amplitudes
are small. This non-superfluid (\textit{i.e} Mott insulating) phase
possesses an XY-ferromagnetic spin-ordering and is characterized by
$\phi_{b,d}=0$ and $\phi_{bd}>0$.

We investigate also the effect of finite temperature on the phase
diagram as shown in the lower panel of Fig.~\ref{fig1new}. We
observe that the superfluid remains robust against small finite $T$.
On the other hand, the XY-ferromagnetic spin-ordered phase is
sensitive to finite temperature since it is formed in the
low-hopping regime and therefore easily destroyed by thermal
fluctuations. At finite $T$ this ordered phase is reduced in favor
of developing a non-magnetically ordered Mott state which is
characterized by $\phi_{bd}=0$ and $\phi_{b,d}=0$ and which we
denote as \textquotedblleft unordered insulator\textquotedblright
$\mbox{}$ in the following. The black dashed line shows the energy
scale of the magnetic coupling $4t_bt_d/U_{bd}$ \cite{Altman2003}.

\begin{figure}
\begin{center}
\includegraphics[clip,width=.9\linewidth]{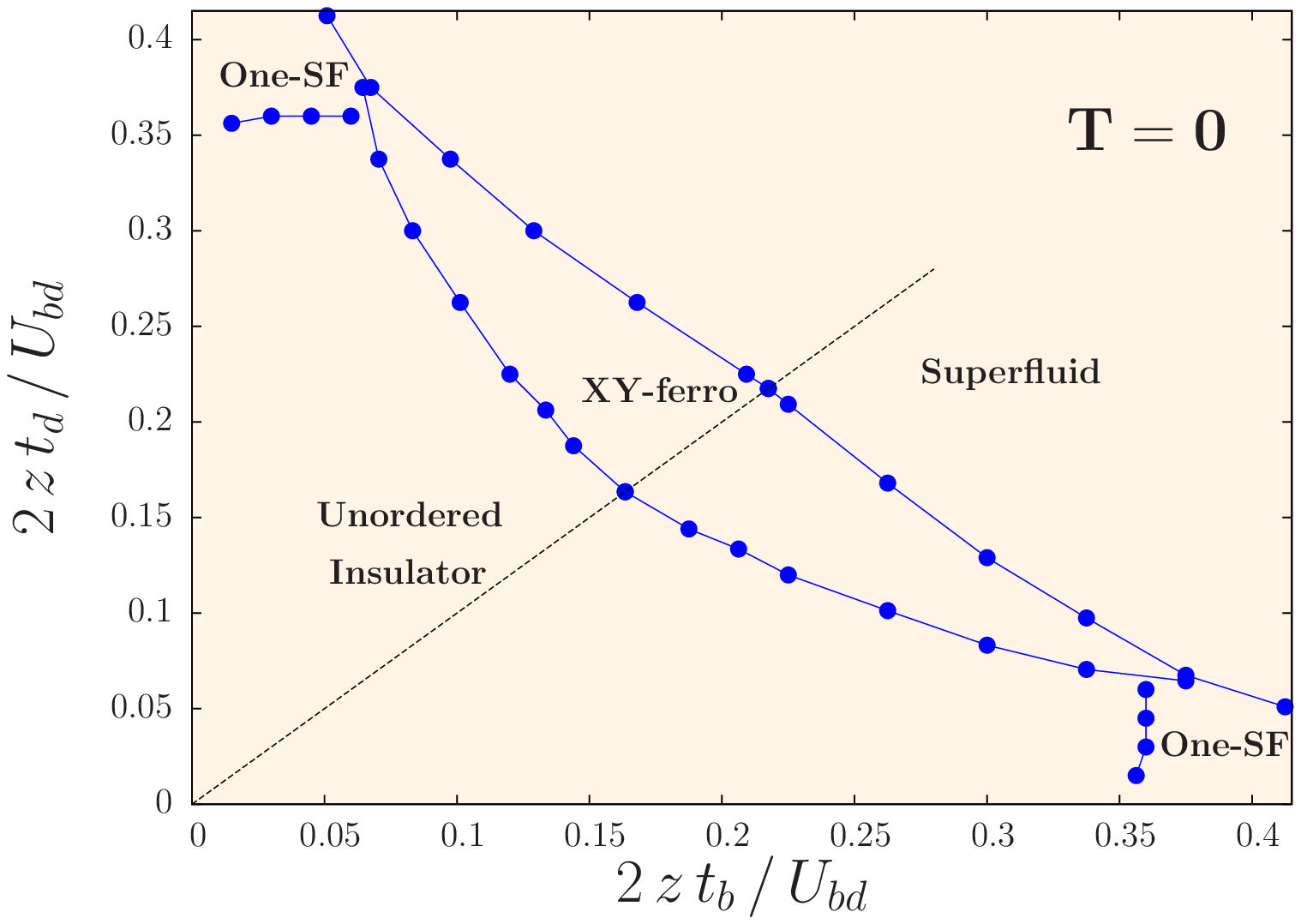}
\includegraphics[clip,width=.9 \linewidth]{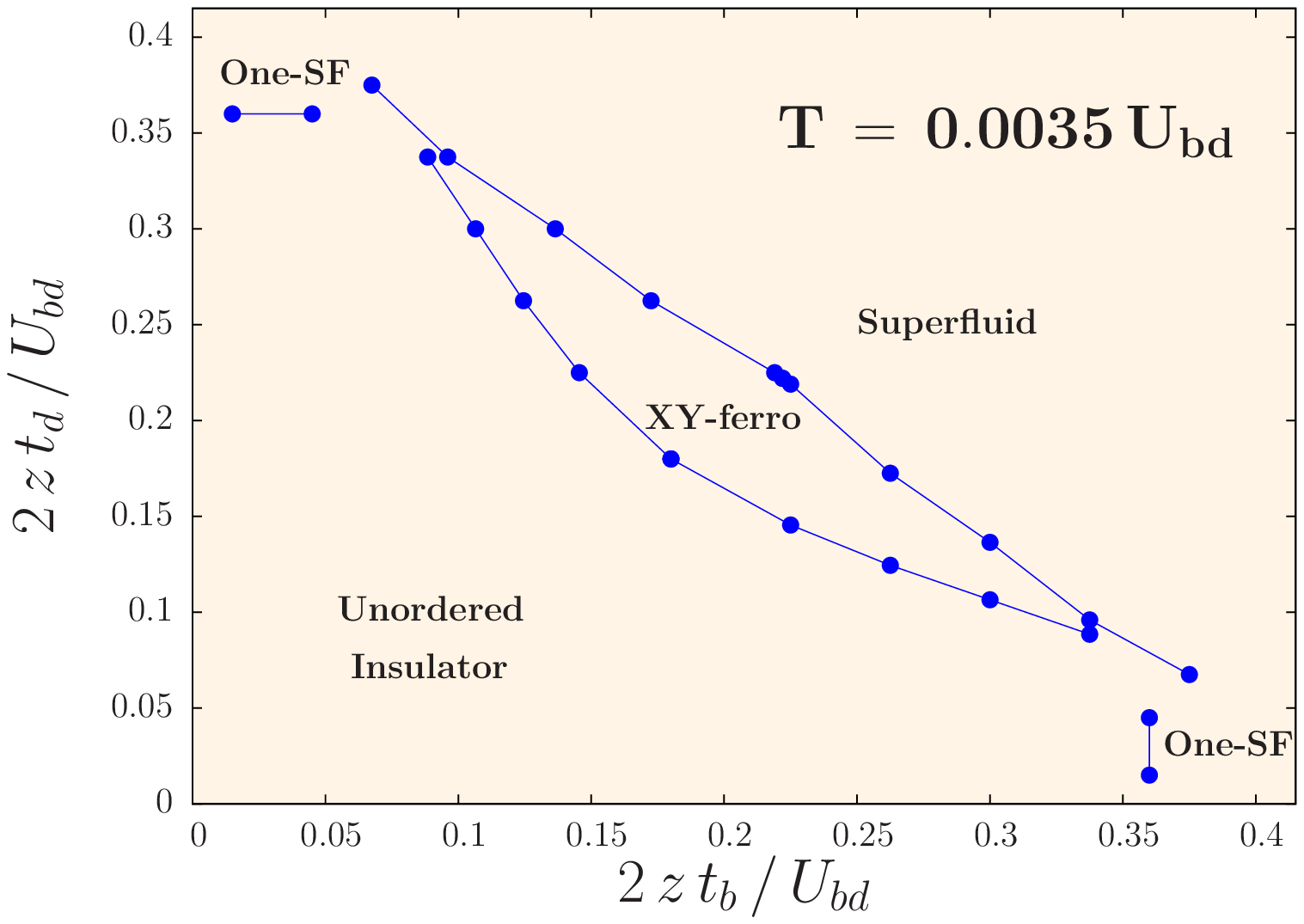}
\caption{ Upper panel: zero-temperature phase diagram for a
two-component Bose-Hubbard model in a  3D cubic lattice, as a
function of hopping parameters. The diagonal dotted line shows
$t_b=t_d$. The interaction values are $U_{b}= U_{d}=1.01\,U_{bd}$
and the total filling is $2$ with $n_b=n_d=1$.
 Lower panel: finite temperature phase diagram ($T = 0.0035 U_{bd}$). } \label{fig2new}
\end{center}
\end{figure}
Now we turn to the case when the total filling at each site is
$n=2$. Fig.~\ref{fig2new} (upper panel) shows the zero-temperature
phase diagram for this case. The main difference compared to $n=1$
is the presence of a large unordered (Mott) insulator at low hopping
values. As for $n=1$, here we also find a sizable superfluid regime
when both species have large hopping amplitudes. Instead, when the
hopping amplitude for one component is very small and the other one
very large, the system will be in a new phase with one component
being superfluid and the other one Mott insulating. This
\emph{one-component superfluid} phase, \textit{e.g.} for the $d$
component, is defined by $\phi_b=0$ and $\phi_d>0$. There are also
two Mott states: the XY-ferromagnet at intermediate hopping and the
unordered (Mott) insulator  in the lower hopping regime. At finite
temperature (lower panel) both superfluid phases remain robust and
almost unchanged. The main effect of finite $T$ is to reduce the
XY-ferromagnetic phase in favor of the unordered Mott insulator. The
finite-$T$ phase diagram of Fig.~\ref{fig2new} is in remarkable
agreement with the one obtained by a field-theoretical approach
\cite{Powell2009}.

\begin{figure}
\begin{center}
\includegraphics[clip,width=.9\linewidth]{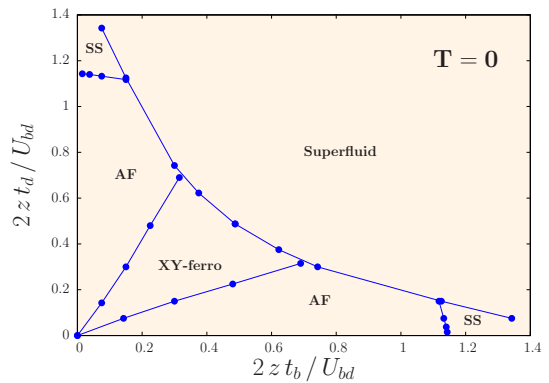}
\caption{ Zero-temperature phase diagram for a two-component
Bose-Hubbard model in a 3D cubic lattice, as a function of hopping
parameters. The interaction values are  $U_{b}= U_{d}=12\,U_{bd}$
and the total filling is $n=1$ with $n_b=n_d=0.5$.} \label{fig3new}
\end{center}
\end{figure}

A further important spin-ordered state is the anti-ferromagnetic
(AF) phase. Its existence in the Mott domain has been shown in
previous investigations \cite{Duan2003,Altman2003}. In order to
address this phase within our formalism, we adopt a set of
parameters in which the inter-species interactions $U_{b,d}$ are
much larger than the intra-species one: $U_b=U_d=12\,U_{bd}$. This
specific choice is inspired by a previous BDMFT study
\cite{Hubener2009} in which the phase diagram was obtained on the
Bethe lattice. Here we aim to map out the phase diagram on a 3D
cubic lattice which is directly relevant for the experimental
studies. Fig.~\ref{fig3new} sketches the phase diagram for this case
with the total particle filling $n=1$. In addition to two previously
discussed phases, superfluid and XY-ferromagnet, (see
Fig.~\ref{fig1new}, \ref{fig2new}), we find two other ordered states
here: AF phase and super-solid. For unequal hopping in the Mott
domain, we identify a magnetically ordered phase of AF type. This
non-superfluid phase (\textit{i.e.} $\phi_{b,d}=0$) is characterized
by a finite value of the AF order parameter
$\Delta_{\text{AF}}^\nu=|n_{\nu,\alpha}-n_{\nu.\bar{\alpha}}|>0$,
where $\nu$ denotes the component and $\alpha$ is the sublattice
$(\bar{\alpha}=-\alpha)$, together with vanishing XY-ferromagnetic
order $\phi_{bd}=0$. Finally for a very large difference in the
hopping of the two species we observe a small window of a
super-solid phase with $\phi_b>0,\phi_d=0$ and
$\Delta_{\text{AF}}>0$ if $t_b\gg t_d$, and vice versa.

\begin{figure}
\includegraphics[clip,width=.75\linewidth]{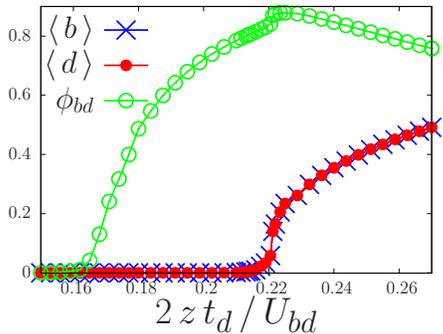}
 \caption{$t_d$ dependence of order parameters $\langle b \rangle$, $\langle d \rangle$,
 and $\phi_{bd}$ along the diagonal black dotted line in the upper panel of Fig. \ref{fig2new}.
 The interaction regime is set to $U_{b}= U_{d}=1.01\,U_{bd}$ and the hopping amplitudes $t_b=t_d$
with filling factors $n_b=n_d=1$.}\label{fig2}
\end{figure}

To investigate in detail the quantum phase transition into the
XY-ferromagnetically ordered state, for $n=2$ and at $T=0$, in
Fig.~\ref{fig2} we plot the behavior of individual superfluid order
parameters and also the correlator $\phi_{bd}$ along the line of
$t_b=t_d$ (the diagonal black dotted line shown in the upper panel
of Fig. \ref{fig2new}). The latter behavior indicates a second-order
quantum phase transition from the Mott state to the XY-ferromagnet
and also a second-order phase transition from XY-ferromagnet to
superfluid.

\begin{figure}[h]
\includegraphics[scale=.6]{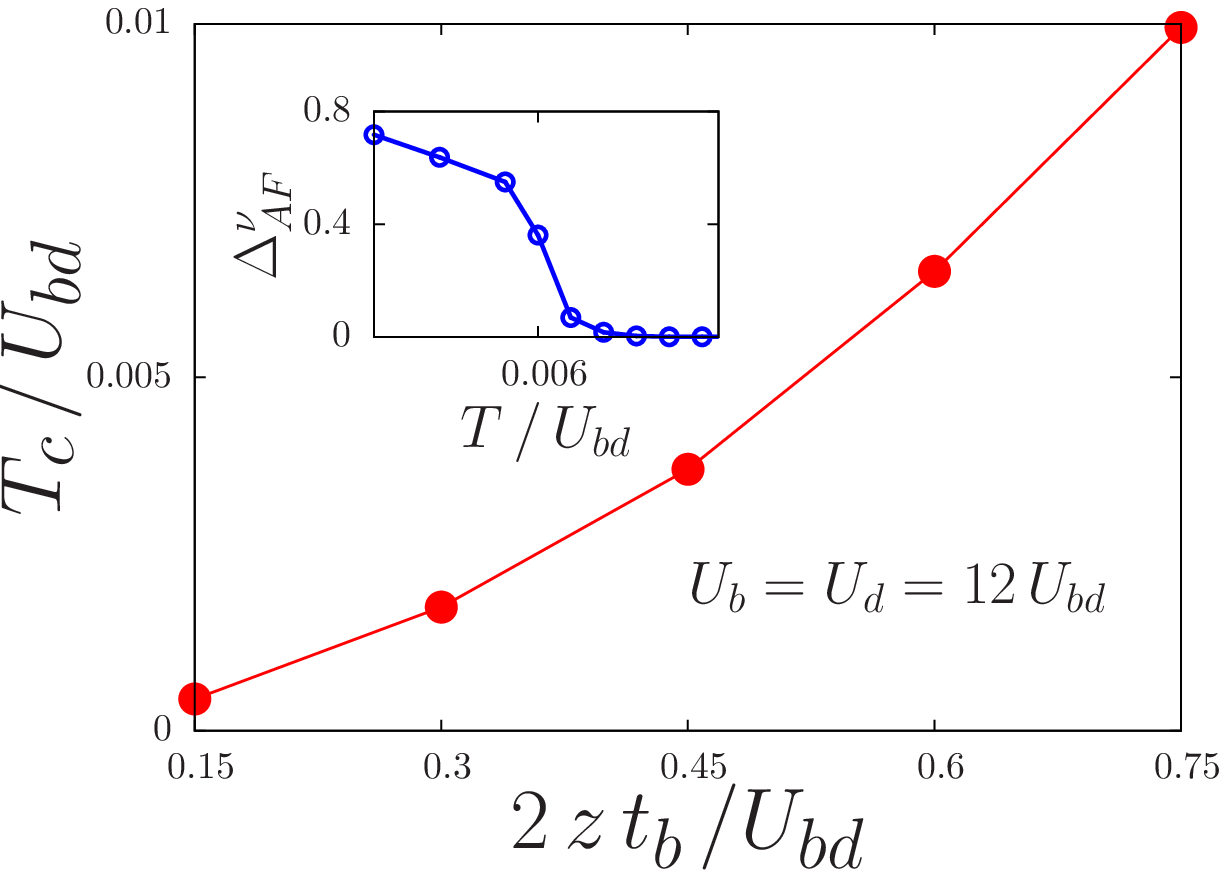}
\includegraphics[scale=.6]{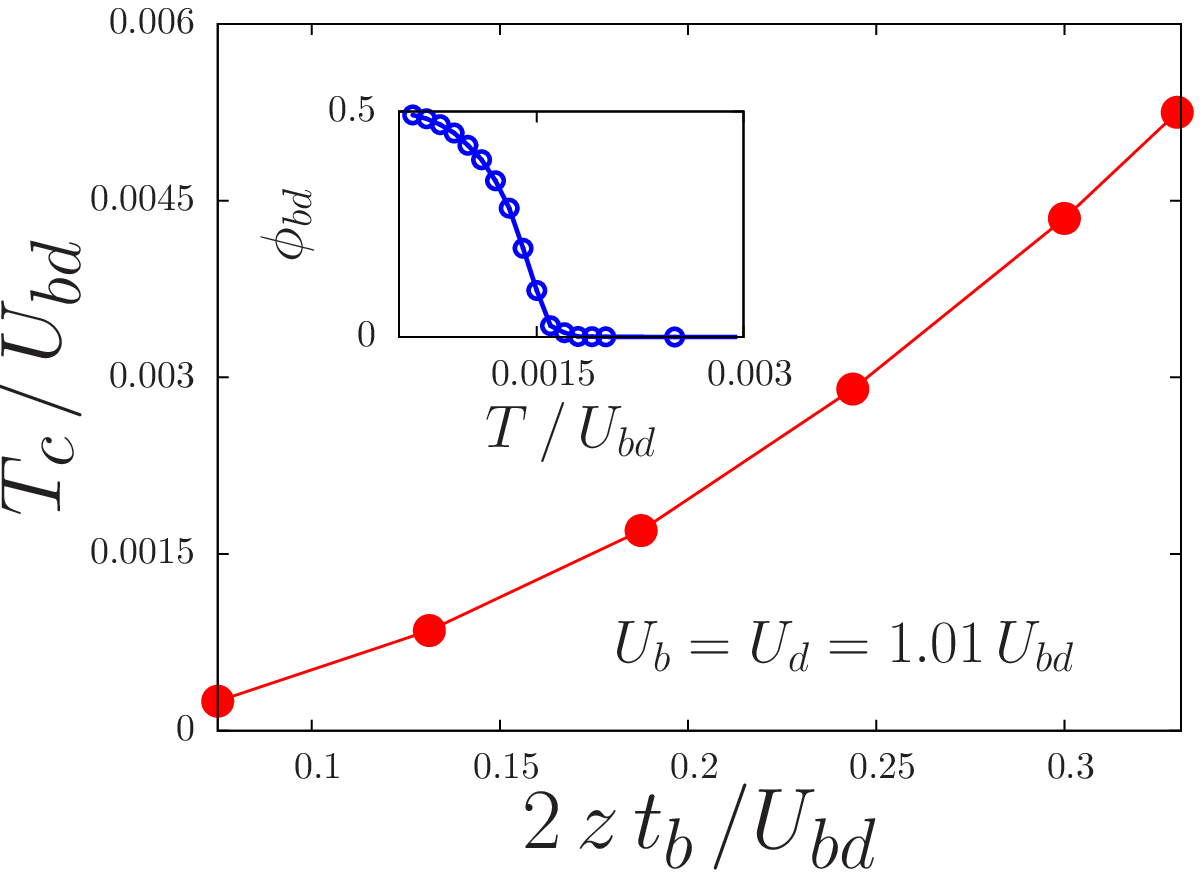}
\caption{Critical temperature of AF and XY-ferromagnetic order as a
function of the hopping amplitude $t_b$ on a 3D homogeneous cubic
lattice with total filling $n=1$. \textbf{Upper panel:} AF phase
with hopping amplitude ratio $t_b=4t_d$. Inset: melting of the AF
phase vs temperature with hopping amplitudes $2zt_{b}=0.6\,U_{bd}$
and $2zt_{d}=0.15\,U_{bd}$. \textbf{Lower panel:} XY-ferromagnetic
phase for equal hopping amplitudes $t_b=t_d$. Inset: melting of the
XY-ferromagnetic phase vs temperature with hopping amplitudes
$2zt_b=2zt_d=0.1875\,U_{bd}$.}\label{fig6}
\end{figure}

One crucial question regarding observation of the AF and
XY-ferromagnetic orders is how fragile they are against
finite-temperature effects. To address this important issue, we
compute the respective critical temperatures. Fig.~\ref{fig6} shows
$T_c$ as a function of the hopping amplitude of species $b$ while we
keep the hopping ratio fixed as $t_d=4t_b$ for the AF phase (upper
panel) and as $t_b=t_d$ for the XY-ferromagnetic phase (lower
panel). We notice that $T_c$ rises as the hopping amplitudes
increase, due to the growing effective exchange couplings which
stabilize magnetic order. We also note that the zero-temperature
phase diagram on a 3D cubic lattice for filling $n=1$, containing
the AF phase (Fig. \ref{fig3new}), has the same structure as the
corresponding one on the Bethe lattice \cite{Hubener2009}. Therefore
we anticipate that the finite-$T$ counterpart of this phase diagram
should also be similar on both lattices, and therefore expect there
is a region of unordered Mott insulator at low hopping. The inset of
Fig.~\ref{fig6} (upper panel) shows the temperature dependence of
$\Delta^\nu_{AF}$. It indicates a second-order phase transition from
the AF phase to the unordered Mott insulator. We have also computed
the order parameter $\phi_{bd}$ for the XY-ferromagnetic phase as
shown in the lower inset of Fig.~\ref{fig6} which indicates a 2nd
order transition from the XY-ferromagnetic phase to an unordered
Mott insulator as well. The critical temperatures of magnetic phases
shown here are notably smaller than the coldest temperatures which
have been measured in most experiments until now, apart from W.
Ketterle's group where temperatures as low as 350 pK ($\approx
0.01U_{bd}$) have been achieved \cite{Medley2010}.

\section{Spin-ordering for an Inhomogeneous Bose-Hubbard model}
In the previous section, we focused on homogeneous systems. However
all the experiments are carried out in the presence of an external
confining potential. Therefore we extend the BDMFT scheme to
real-space BDMFT to address the inhomogeneous system. In this
section, using RBDMFT and LDA+BDMFT, we will explore an
inhomogeneous two-component Bose-Hubbard model both in 2D and 3D.

\subsection{2D trapped Bose gas}
In this section we discuss the AF phase, the XY-ferromagnet and the
unordered Mott state in a 2D square lattice in the presence of a
harmonic trap. We first investigate AF ordering on a $31\times31$
lattice and then XY-ferromagnetic phase and the unordered Mott state
on a $32\times32$ lattice. The choice of different lattice sizes is
solely due to computational convenience.

\subsubsection{\textbf{Anti-ferromagnet order}}
\begin{figure}
\begin{center}
\includegraphics[clip,width=.6\linewidth]{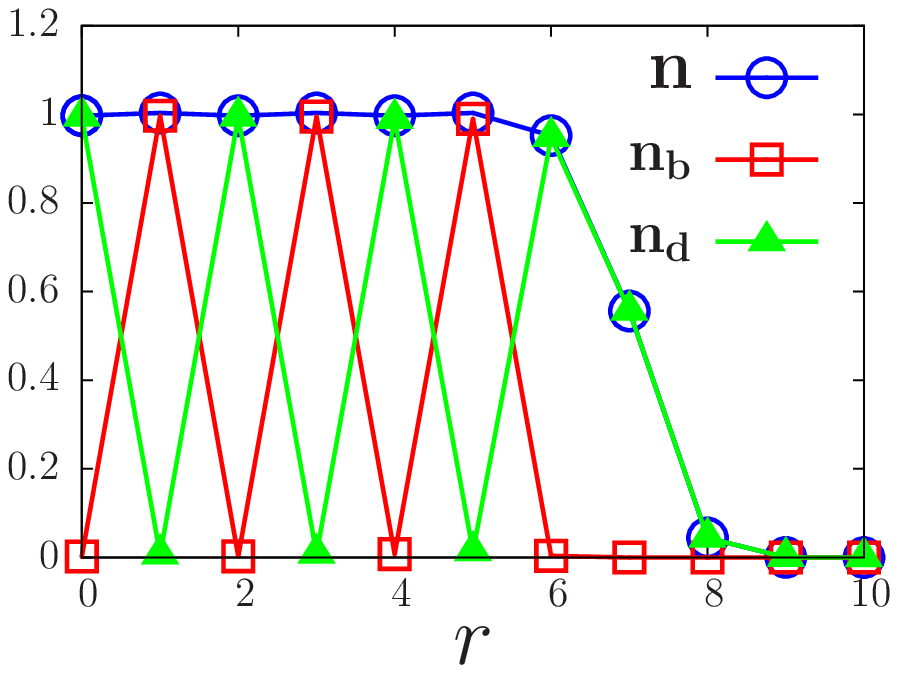}
\includegraphics[clip,width=.55\linewidth]{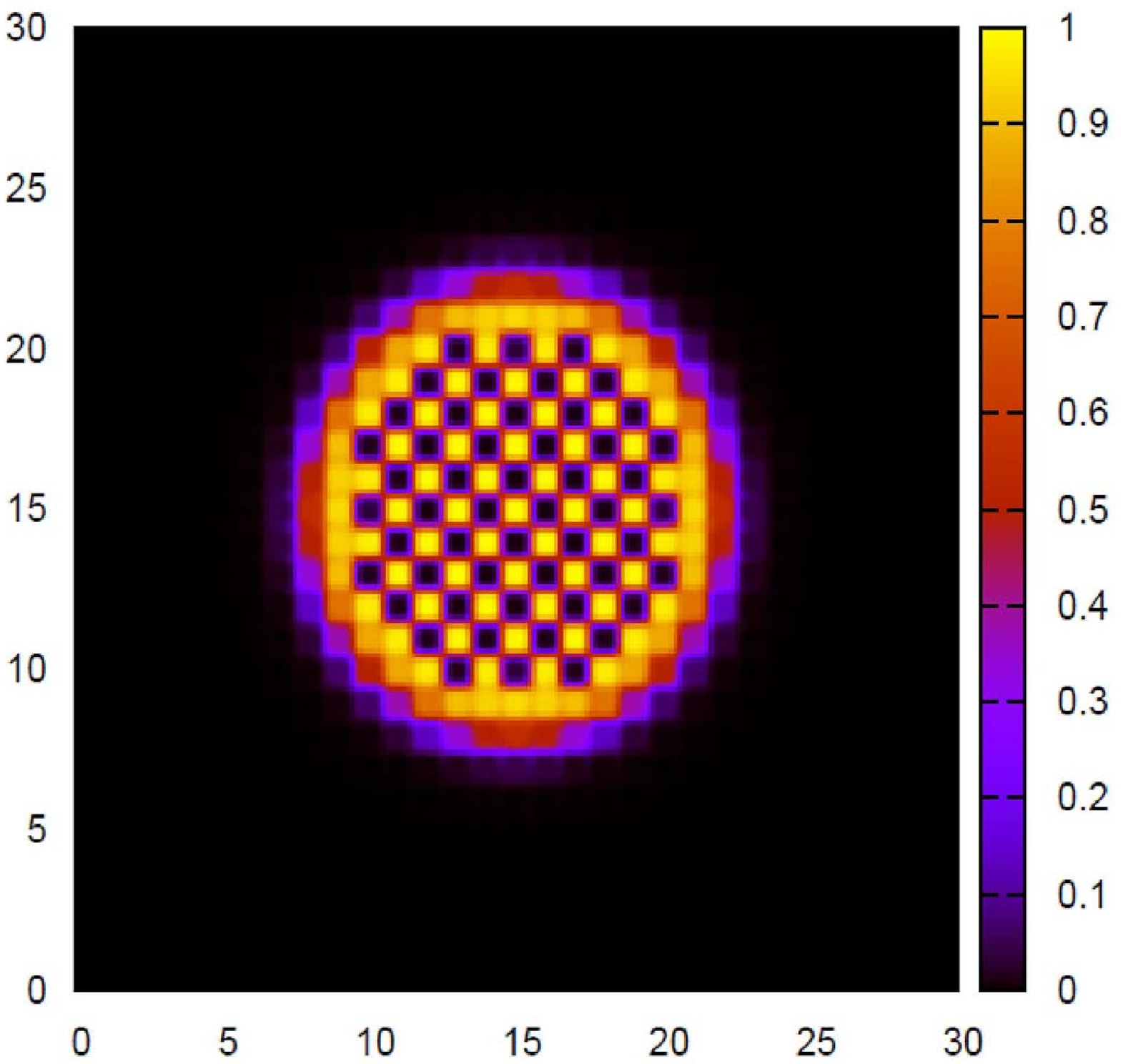}
\caption{ N\'eel-type AF order in 2D at central filling $n=1$ and
$T=0$. \textbf{Top:} particle densities of the two species as a
function of radial distance $r$. The interactions are set to $U_{b}=
U_{d}=12\,U_{bd}$, the hopping amplitudes $2zt_b=0.1\,U_{bd}$ and
$2zt_d=0.25\,U_{bd}$ with harmonic trap $V_{0}=0.01\,U_{bd}$. The
chemical potentials are $\mu_b=\mu_d=0.5\,U_{bd}$. \textbf{Bottom:}
density distribution of the $d$ component over the lattice.
}\label{fig4}
\end{center}
\end{figure}
\begin{figure}
\begin{center}
\includegraphics[clip,width=.6\linewidth]{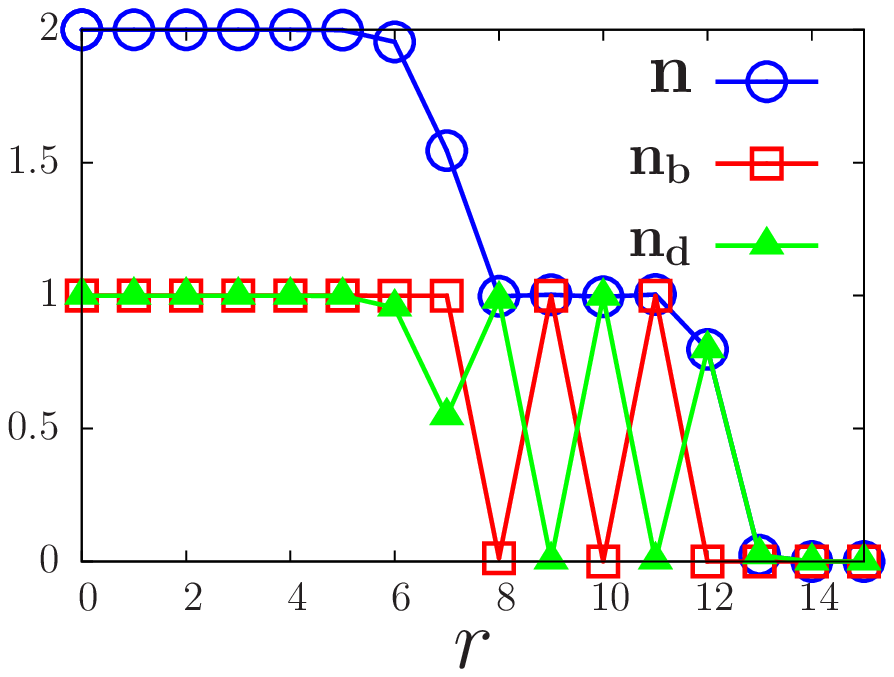}
\includegraphics[clip,width=.55\linewidth]{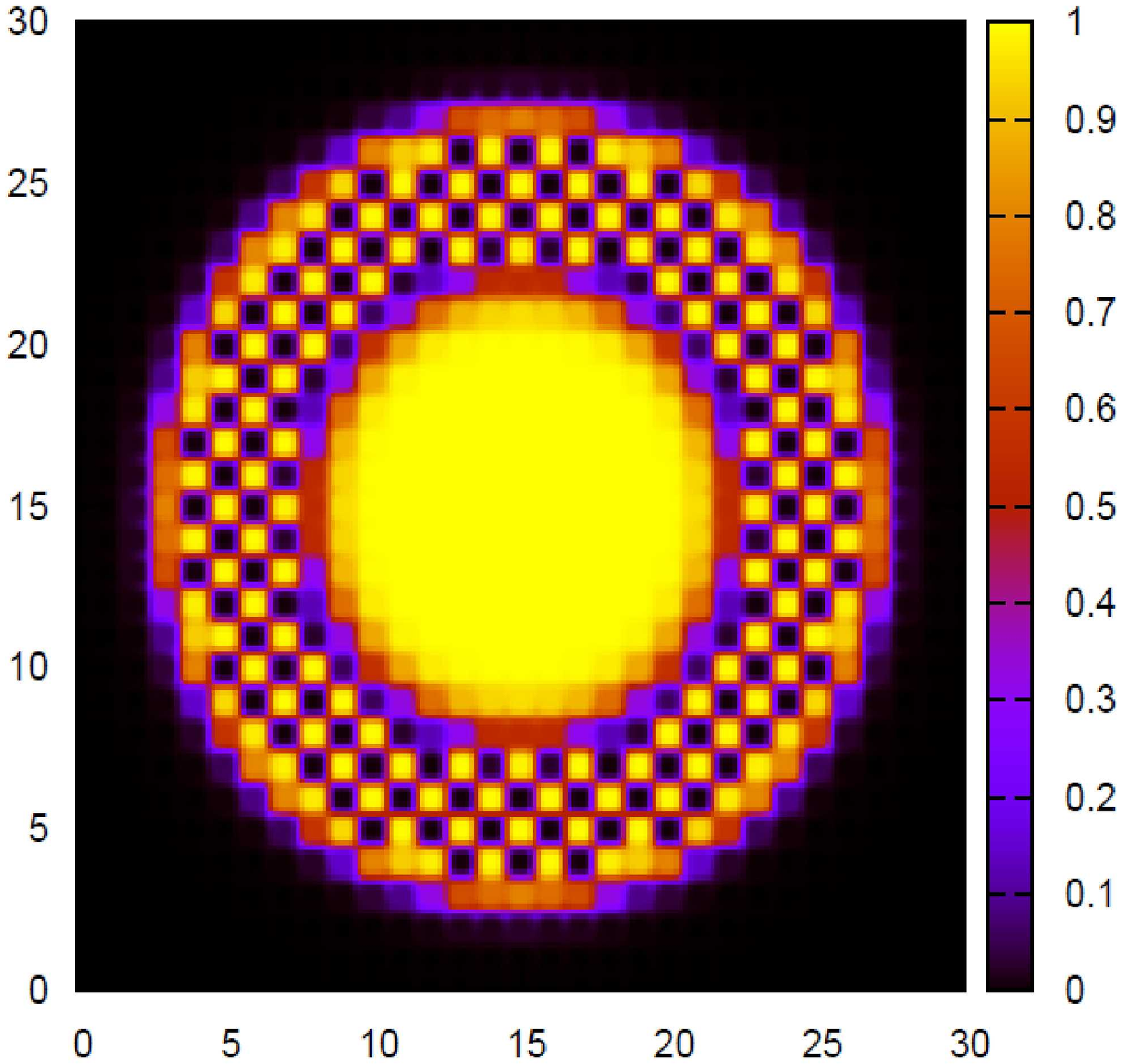}
\caption{ N\'eel-type AF order in 2D at central filling $n=2$ and
$T=0$. \textbf{Top:} particle densities of the two species versus
radius obtained by RBDMFT. Interactions are set to $U_{b}=
U_{d}=48\,U_{bd}$, the hopping amplitudes to $2zt_b=0.1\,U_{bd}$ and
$2zt_d=0.25\,U_{bd}$, with a harmonic trap strength
$V_{0}=0.01\,U_{bd}$. The chemical potentials are
$\mu_b=\mu_d=1.5\,U_{bd}$. \textbf{Bottom:} density distribution of
the $d$ component over the lattice.}\label{fig5}
\end{center}
\end{figure}

One of the most desirable goals in the current experiments on cold
atomic-gases is to reach the regime of (N\'eel-type) AF ordering
which is (for fermionic systems) expected to be a key step towards
realizing a $d$-wave superfluid phase \cite{Esslinger2010}. Here we
investigate this phase for trapped two component bosons in an
optical lattice. At the beginning we focus on AF order at $T=0$. We
choose the interaction parameters as $U_b=U_d=12\,U_{bd}$ with the
unequal hopping for two species fixed as $2zt_b=0.1\,U_{bd}$ and
$2zt_d=0.25\,U_{bd}$. We choose a maximum local filling of $n=1$ at
the center of the trap. The top panel of Fig.~\ref{fig4} shows the
RBDMFT results for the particle densities as a function of radial
distance $r$ from the trap center. The AF phase forms in the central
area of the lattice as a checker-board pattern  and vanishes
smoothly with increasing distance $r$ from the lattice center. This
indicates that AF order is stable at the center of a finite trap.
However due to the unequal hopping, the lighter species
(\textit{i.e.} the one with the larger hopping) explores the lattice
more freely and forms a superfluid ring around the central
checker-board pattern. This behavior is visible in the bottom panel
of Fig.~\ref{fig4}.  To see how robust AF order is against changing
the total atom filling number, we now increase the filling at the
center of the trap to $n=2$, and observe that AF order now forms as
a ring around the center as shown in Fig.~\ref{fig5}. We conclude
that also in 2D and in the presence of a trapping potential, AF
order can be observed in regions of total filling $n=1$ at $T=0$.

\subsubsection{\textbf{XY-ferromagnet}}\label{XY-ferromagnet}
\begin{figure}
\begin{center}
\includegraphics[clip,width=.7\linewidth]{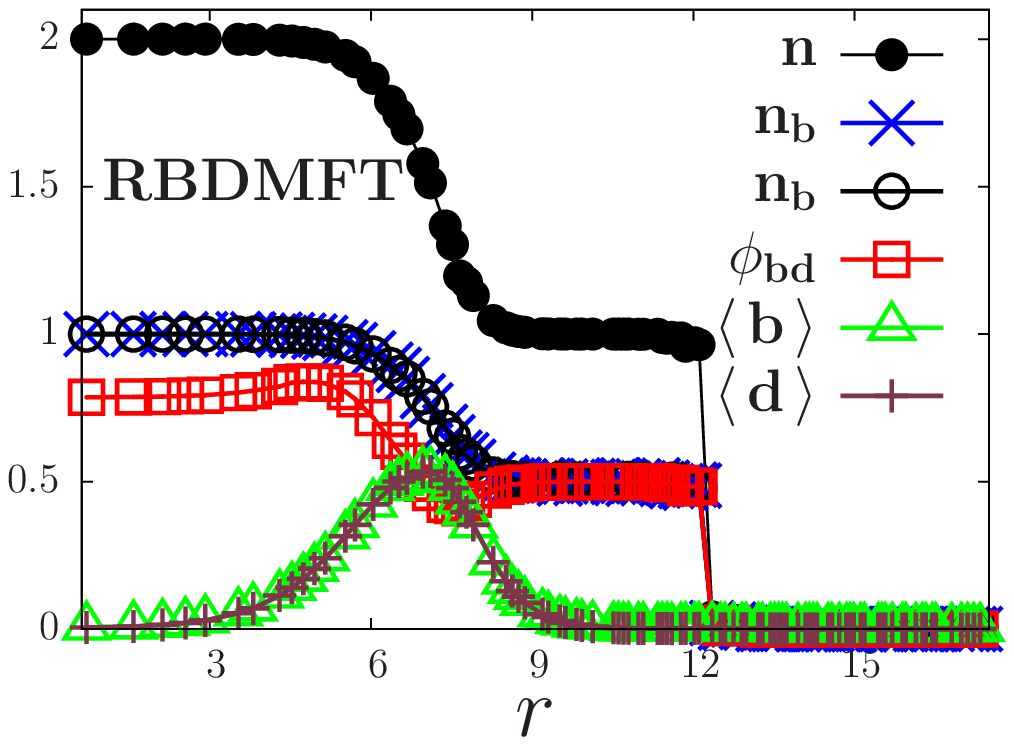}
\includegraphics[clip,width=.7\linewidth]{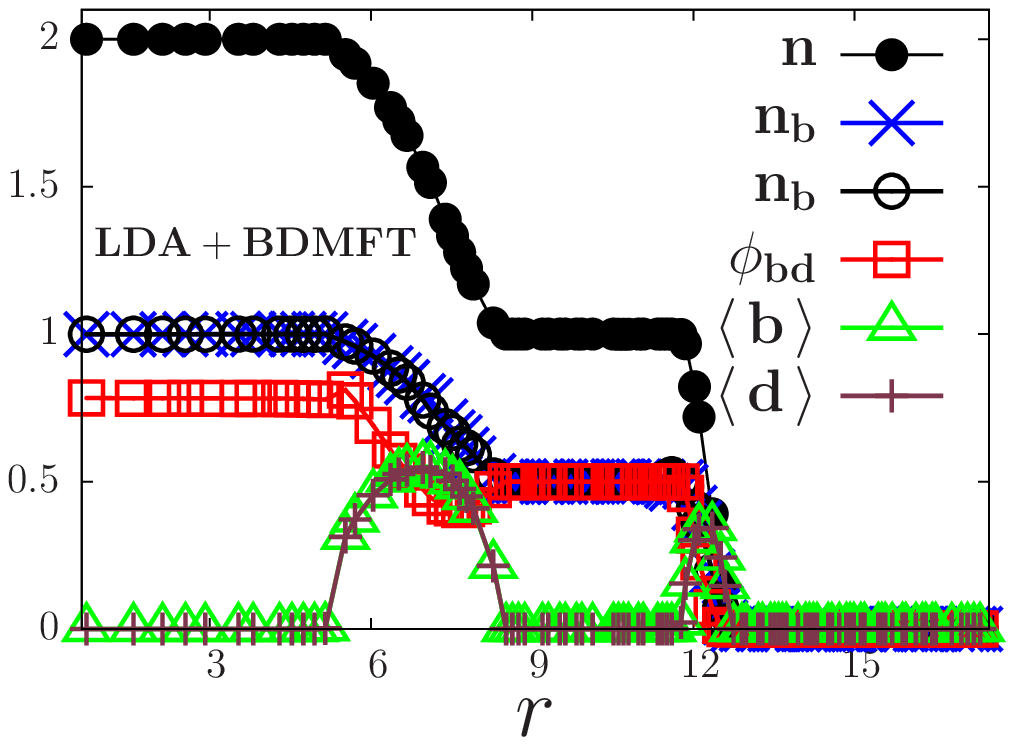}
\caption{Top: RBDMFT results in 2D for atom densities, superfluid
order parameters and XY-ferromagnetic correlator $\phi_{bd}$ as a
function of radial distance $r$ at $T=0$. Interactions are set to
$U_{b}= U_{d}=1.01\,U_{bd}$, with hopping amplitudes
$2zt_b=2zt_d=0.175\,U_{bd}$ and harmonic trap $V_{0}=0.01\,U_{bd}$.
The chemical potentials are $\mu_b=\mu_d=1.5\,U_{bd}$. Bottom:
LDA+BDMFT results for the same parameters as the top
panel.}\label{fig7}
\end{center}
 \end{figure}
As evident in the phase diagrams of the homogeneous system
(Figs.~\ref{fig1new}, \ref{fig2new}), a common magnetic phase which
appears for both fillings $n=1$ and $n=2$ is the XY-ferromagnet.
Here we investigate the stability of the XY-ferromagnetic phase in a
trapped 2-component system on a 2D lattice of size $32\times32$ at
$T=0$. We first focus on the case of equal hopping
$2zt_b=2zt_d=0.175\,U_{bd}$  for both species and choose the
interactions as $U_b=U_d=1.01U_{bd}$. In Fig. \ref{fig7} the atom
densities, their corresponding superfluid order parameters and the
correlator $\phi_{bd}$ are shown as determined from RBDMFT (top
panel) and LDA+BDMFT (lower panel). At the center of the lattice, we
have a total filling of $n=2$. We observe a finite value of the
correlator $\phi_{bd}$ which implies a stable XY-ferromagnetic phase
in this inhomogeneous system. With increasing distance from the trap
center, we find non-zero values for the superfluid order parameters
with a maximum inside the atomic cloud, indicating the superfluidity
for both species. We can also see that the correlator $\phi_{bd}$
remains finite in the superfluid regime. Moving further towards the
edge of the trap, the XY-ferromagnetic phase with $n=1$ appears and
eventually a further superfluid domain. For comparison, we have
computed the same quantities within LDA+BDMFT, which are shown in
the bottom panel of Fig.~\ref{fig7}. We find a excellent agreement
between RBDMFT and LDA+BDMFT deep inside each phase, but RBDMFT
provides the more accurate description of the smooth transition
between the different phases. Note that the second superfluid ring
is very narrow within RBDMFT which is most likely a finite-size
effect.

\subsubsection{\textbf{Unordered Mott state}}
\begin{figure}
\begin{center}
\includegraphics[clip,width=.75\linewidth]{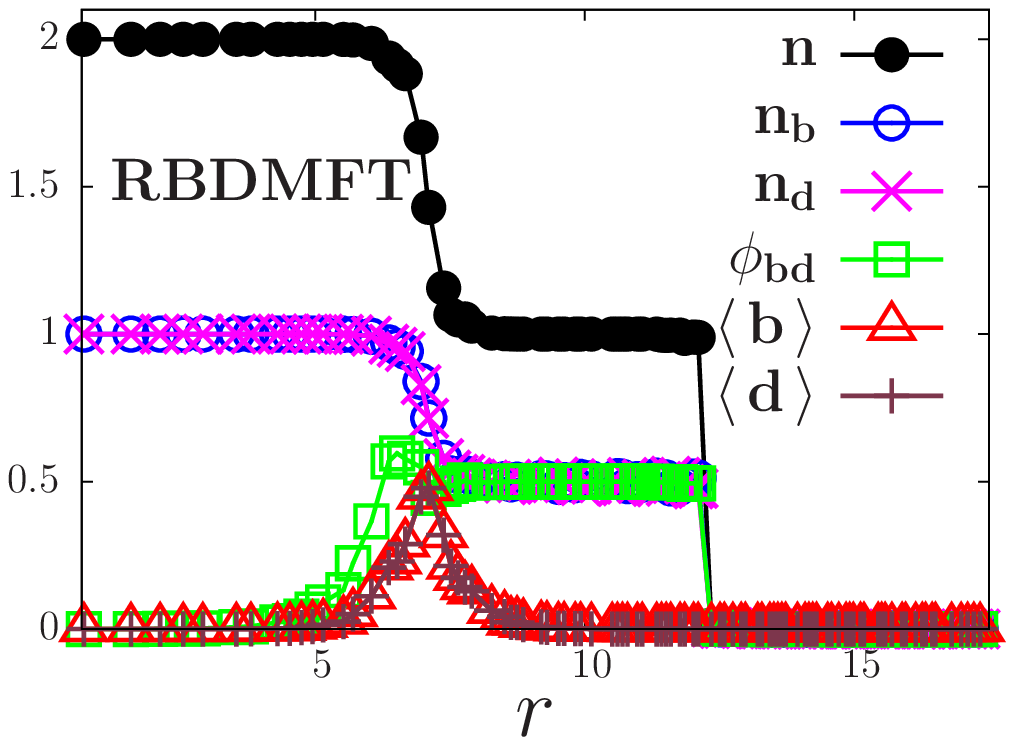}
\includegraphics[clip,width=.75\linewidth]{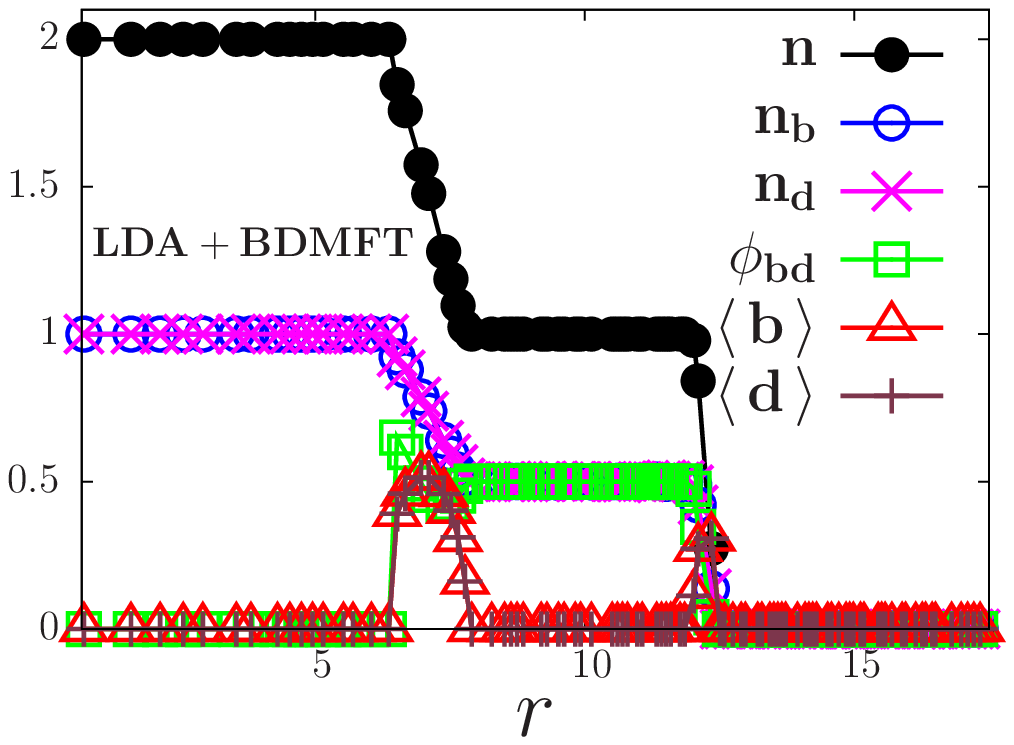}
 \caption{Unordered Mott state in 2D at $n=2$ for weak hopping amplitudes at $T=0$,
 calculated by RBDMFT (top panel) and LDA+BDMFT (bottom panel). The interactions and hopping
 amplitudes are $U_{b}= U_{d}=1.01\,U_{bd}$ and  $2zt_b=2zt_d=0.1\,U_{bd}$, with the harmonic trap $V_0=0.01\,U_{bd}$.
 The chemical potentials are $\mu_b=\mu_d=1.5\,U_{bd}$.}\label{fig11}
\end{center}
\end{figure}

We now consider the case of low hopping amplitudes. This situation
corresponds to the regime of the large area of the unordered Mott
state (without symmetry breaking) in the homogeneous phase diagram
(see Figs.~\ref{fig1new},\ref{fig2new} for the 3D case). In
Fig.~\ref{fig11} we show results for the atomic densities,
superfluid order parameters, and the correlator $\phi_{bd}$ for the
case $2zt_b=2zt_d=0.1\,U_{bd}$ at $T=0$. In the center of the
lattice where we have filling $n=2$, the low-energy Hilbert space of
each lattice site includes the three states $|b,b\rangle ,
|d,d\rangle$ and $|b,d\rangle$. With the choice of the interactions
as $U_{b,d}>U_{bd}$, the third state has the lowest energy and
therefore we have a Mott state with $\phi_{bd}=0$ and no symmetry
breaking. For intermediate radii, where $1<n<2$ and both species are
superfluid, $\phi_{bd}$ rises to a finite value and remains constant
in the $n=1$ region where the XY-ferromagnet is stable.

\subsection{3D trapped case}
In this final part, we consider the experimentally most interesting
case of a 3D cubic optical lattice in the presence of an external
harmonic trap. Due to the computational limitations for RBDMFT, here
we only apply LDA+BDMFT which we previously benchmarked versus
RBDMFT. Throughout this section, we consider a lattice with
$41\times 41 \times 41$ sites.

\subsubsection{\textbf{XY-ferromagnet}}
\begin{figure}
\begin{center}
\includegraphics[clip,width=.75\linewidth]{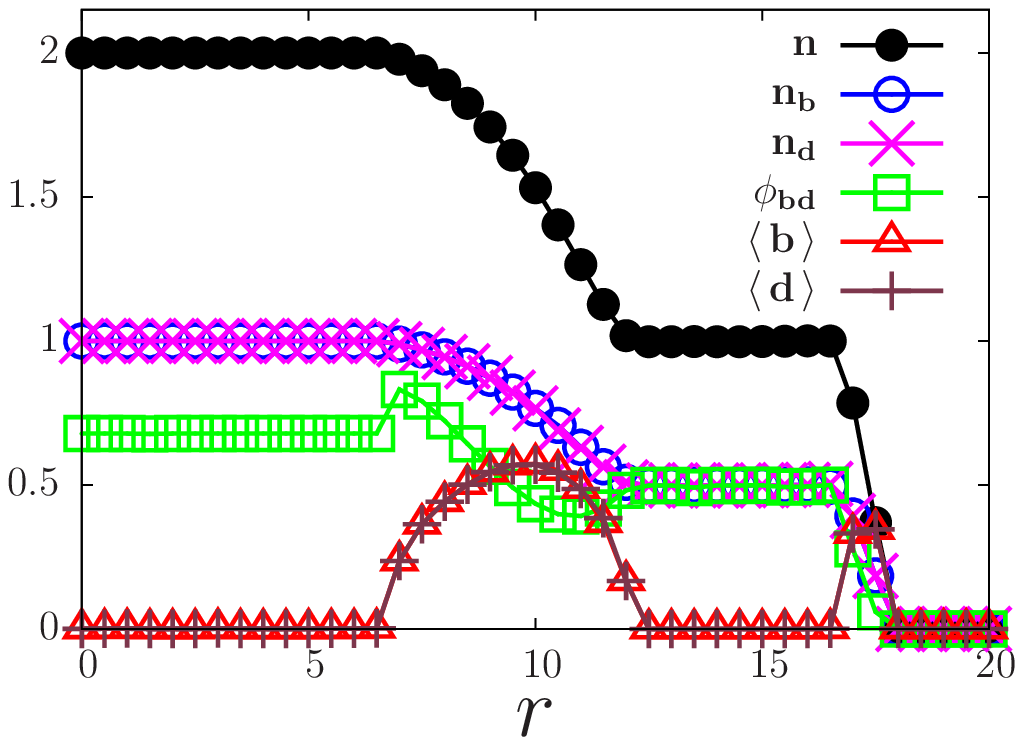}
\includegraphics[clip,width=.78\linewidth]{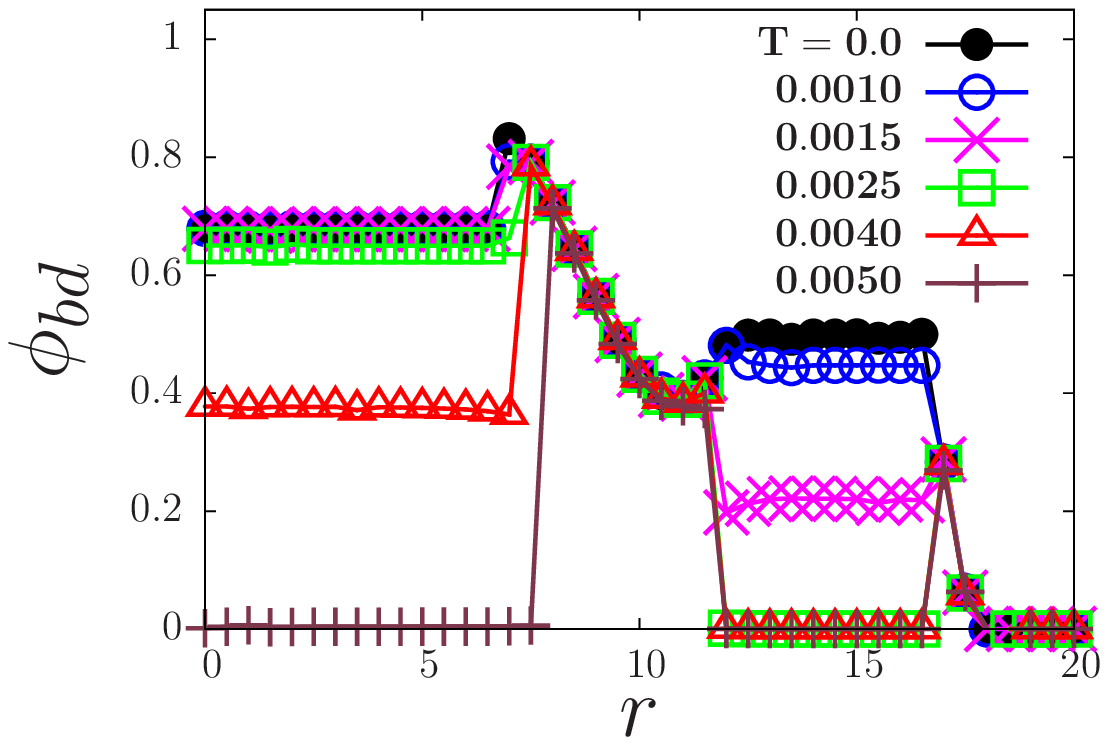}
\caption{Top: particle densities, superfluid order parameters and
XY-ferromagnetic correlator as a function of radius $r$ on a 3D
cubic lattice at $T=0$, calculated within LDA+BDMFT. Interactions
are set to $U_{b}= U_{d}=1.01\,U_{bd}$, hopping amplitudes
$2zt_b=2zt_d=0.195\,U_{bd}$ with a harmonic trap
$V_{0}=0.005\,U_{bd}$ and $N_{tot}=2.6\times 10^4$. Bottom:
Temperature dependence of the XY-ferromagnetic correlator for the
same parameters as the top panel.} \label{fig13}
\end{center}
\end{figure}
We begin the 3D trapped lattice analysis by investigating the
stability of the XY-ferromagnetic phase. We first choose
intermediate hopping as $2zt_b=2zt_d=0.195\,U_{bd}$. This choice
corresponds to the XY-ferromagnet in the homogeneous phase diagrams
(Fig.~\ref{fig1new},\ref{fig2new}). We enforce filling  $n=2$ at the
center of the lattice by adopting the proper chemical potentials. In
Fig.~\ref{fig13} (top panel) we show the particle densities,
superfluid order parameters and the correlator $\phi_{bd}$ at $T=0$
(top panel) and at finite $T$ (bottom panel). For $T=0$ we observe a
wedding-cake structure of the atomic densities \textit{i.e.} two
plateaus of $n=1$ and $n=2$ and intermediate areas with non-integer
filling. However we are more interested in the magnetic properties
which are revealed by the correlator $\phi_{bd}$. We observe that in
the $n=2$ domain, there is a XY-ferromagnetic phase, manifested by a
finite value of $\phi_{bd}$ and vanishing superfluid order
parameters. In the intermediate area $1<n<2$, we obtain a superfluid
phase with both $\langle b \rangle$ and $\langle d \rangle$ being
finite. Note that the onset of superfluidity leads to non-zero
XY-ferromagnetic correlations as well. By approaching the second
Mott plateau with $n=1$, the superfluid order parameters vanish
again and we obtain a non-zero value for $\phi_{bd}$, indicating
once again an insulating XY-ferromagnet. Finally for $n<1$ we find a
further superfluid domain.

As in the previous section, we are interested in the effect of
temperature on magnetic order. Fig.~\ref{fig13} (bottom panel)
represents the correlator $\phi_{bd}$ for different temperatures.
First we notice that for all the temperatures considered here, the
correlator possesses a larger value at $n=2$ compared to $n=1$. In
other words, XY-ferromagnetic order is more pronounced for $n=2$
compared to $n=1$ as long as all other parameters of the
Bose-Hubbard model are identical. To make this point more clear, we
calculate the critical temperature for ferromagnetic order for both
$n=1$ and $n=2$, and find respectively $T_c=0.0018 \, U_{bd}$
($\approx 70$ pK) and $T_c=0.0051\,U_{bd}$ ($\approx 190$ nK). We
also calculate the maximum value of the critical temperature for
XY-ferromagnetic order at filling $n=1$ and $n=2$, and find that the
maximum value of the critical temperature is around 200 pK and 230
pK, respectively, when the 3D cubic lattice is formed by laser beams
of wave-length $1064$ nm and the scattering length is around
$100a_b$ ($a_b$ is the Bohr radius). This fact could be significant
for ongoing experiments, \textit{e.g.} in Refs.
\cite{Weld2009,Medley2010} where spin gradient thermometry has been
used to measure temperatures as low as 350 pK in a 3D optical
lattice. Our calculation here indicates that it is much easier to
observe XY-ferromagnetism for higher filling due to the enhanced
critical temperature.

\begin{figure}[h]
\begin{tabular}{ cc }
\includegraphics[scale=.475] {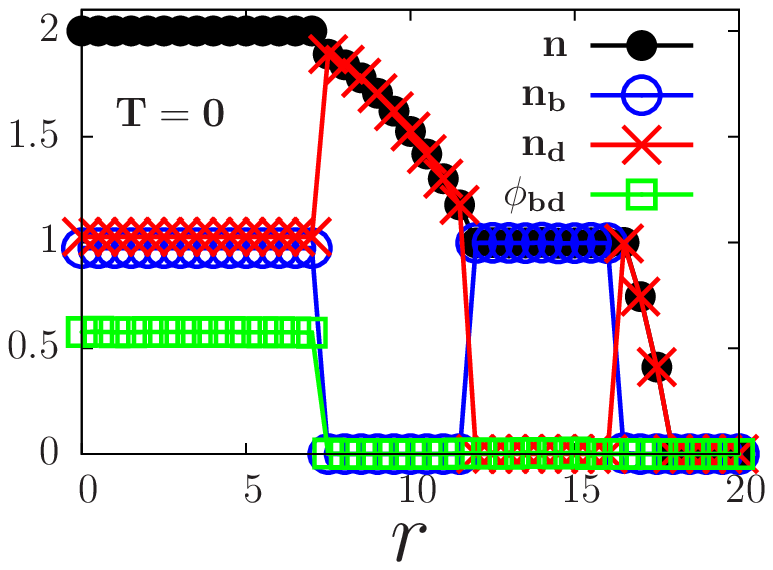}&
\includegraphics[scale=.475]{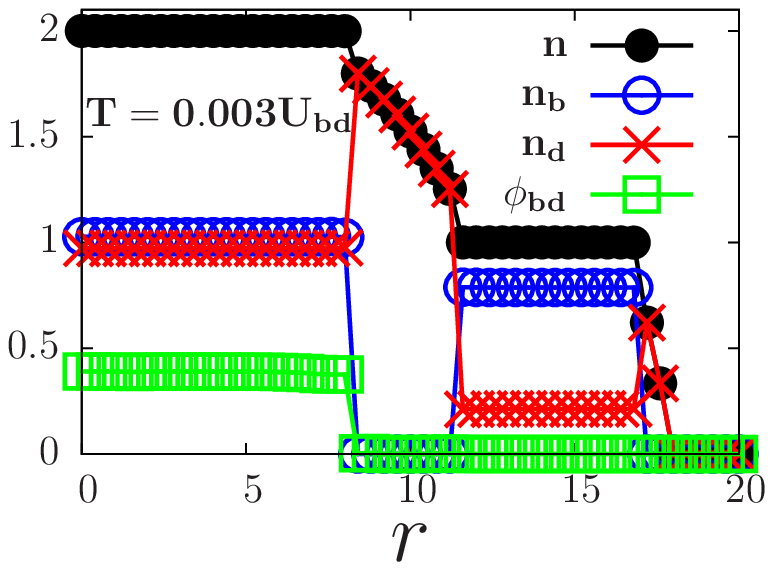}
\\
\includegraphics[scale=.475] {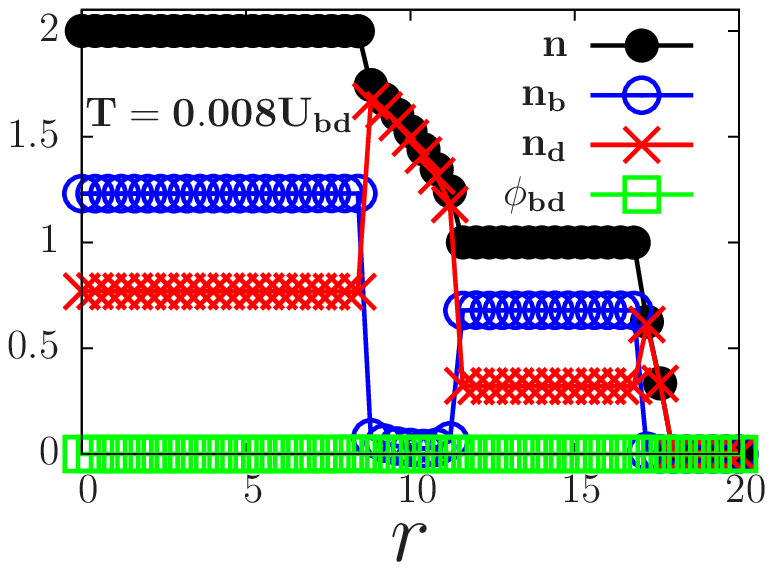}&
\includegraphics[scale=.475]{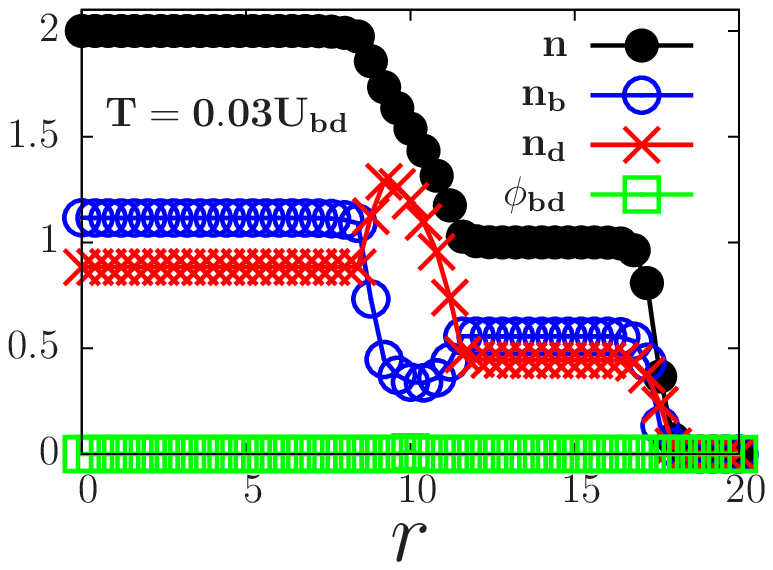}
\end{tabular}
\caption{Particle densities and XY-ferromagnetic correlator vs
radial distance $r$ for asymmetric hopping amplitudes
$2zt_b=0.15\,U_{bd}$ and $2zt_d=0.225\,U_{bd}$ on a 3D cubic
lattice. Interactions are $U_{b}= U_{d}=1.01\,U_{bd}$, the trapping
potential is $V_{0}=0.005\,U_{bd}$ and $N_{tot}=2.6\times
10^4$.}\label{fig14}
\end{figure}
We now consider asymmetric values for the hopping amplitudes
$2zt_b=0.15 \,U_{bd}$ and $2zt_d=0.225 \,U_{bd}$. By adjusting the
chemical potentials, we obtain a globally almost balanced mixture
with $N_b\simeq N_d\simeq (48\%-52\%)\,N_{tot}$ with
$N_{tot}=2.6\times 10^4$.  Fig.~\ref{fig14} shows the atomic
densities and XY-ferromagnetic spin-order for different temperatures
(four panels). The most remarkable new feature of the asymmetric
hopping regime is the vanishing ferromagnetic order in the $n=1$
domain where $\phi_{bd} =0$ even at $T=0$. We also find that the
total density profile first becomes sharper and then smoother again
with increase of temperature. This is due to the higher spin entropy
of the unordered two-component Mott insulator.
Details will be discussed in a future publication \cite{entropy}.

\begin{figure}
\begin{center}
\includegraphics[clip,width=.75\linewidth]{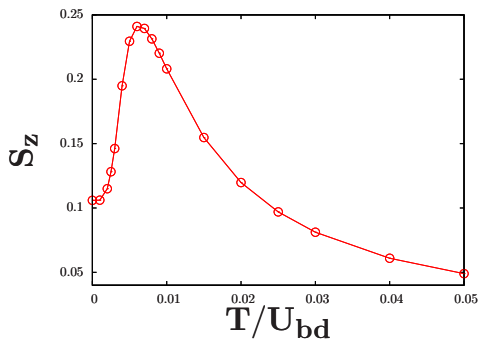}
\caption{Z-magnetization (imbalance) $S_z$ vs temperature for the
\textit{homogeneous} system. Interactions are set to $U_{b}=
U_{d}=1.01\,U_{bd}$ with hopping amplitudes $2zt_b=0.15 \,U_{bd}$
and $2zt_d=0.225\,U_{bd}$ and total filling $n=2$. }\label{fig15}
\end{center}
\end{figure}

It is visible in Fig.~\ref{fig14} that the lighter species
(\textit{i.e.} the one with larger hopping) always dominates the
density distribution in the superfluid area where $n\neq 1,2$, since
in this regime the particle mobility plays a more important role
than in the Mott domains. For the Mott state ($n=1,2$) the situation
differs. In the spin model language
\cite{Kuklov2003,Duan2003,Altman2003}, different hopping amplitudes
lead to the existence of an effective magnetic field
\cite{Kuklov2003,Altman2003},
\begin{equation*}
 h=z(2S+1)\,\frac{t_b^2-t_d^2}{U_{bd}}+\mu_b-\mu_d.
\end{equation*}
where $S=n/2$. This effective magnetic field gives rise to an
imbalance between $n_b$ and $n_d$ which we quantify by the
Z-magnetization $S_z=(n_b-n_d)/2$. At the Mott plateau with $n=1$,
the magnetization will shrink from $1/2$ (maximum of Z-magnetization
due to a large effective magnetic field) to zero with increasing $T$
due to thermal fluctuations.
At the Mott plateau with $n=2$, the magnetization depends
non-monotonically on temperature. In Fig.~\ref{fig15}, we show the
temperature dependence of the magnetization at filling $n=2$ plateau
by focusing on the trap center and performing a finite-$T$ study
with a single-site BDMFT for the \textit{homogeneous } model. This
behavior can be understood if we notice that in the filling $n=2$
region the system favors a state with a small magnetization $S_z$ at
zero temperature due to a non-zero effective magnetic field. Once
the temperature starts to increase from zero to a finite value,
thermal fluctuations will come into play and compete with quantum
fluctuations, which makes the imbalance reach a maximum at finite T,
since both types of fluctuations can delocalize the atoms and smooth
the imbalance between the two species. When the temperature
increases even higher, the larger thermal fluctuations will simply
smear out the imbalance.

\subsubsection{\textbf{Unordered Mott state}}
\begin{figure}[h]
\begin{center}
\includegraphics[clip,width=.75\linewidth]{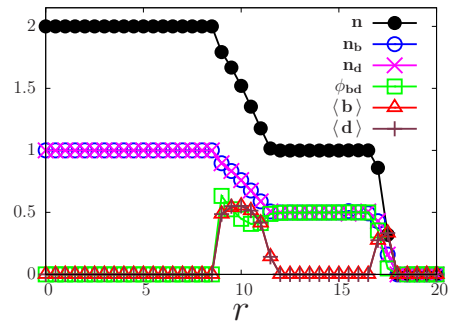}
\caption{Unordered Mott state at $n=2$ for weak hopping amplitudes
at $T=0$, calculated within LDA+BDMFT. Interactions are $U_{b}=
U_{d}=1.01\,U_{bd}$, hopping amplitudes $2zt_b=2zt_d=0.15 \,U_{bd}$,
with a harmonic trap $V_{0}=0.005\,U_{bd}$ and $N_{tot}=2.6\times
10^4$. } \label{fig16}
\end{center}
\end{figure}

Finally, we also investigate the non-magnetic Mott state without
symmetry breaking which occurs at $n=2$ for relatively low hopping
amplitudes $2zt_b=2zt_d=0.15\,U_{bd}$ (see also the homogeneous
phase diagrams Fig.~\ref{fig2new}). Fig.~\ref{fig16} (top panel)
shows the results for atomic densities, their superfluid order
parameters and the correlator $\phi_{bd}$. At the center of the trap
we indeed find a Mott state without symmetry breaking similar to the
2D case (Fig.~\ref{fig11}).

\section{Summary and Outlook}

In conclusion, we have studied magnetic ordering of a two-component
Bose gas in 2D and 3D optical lattices. By using BDMFT we have
calculated the phase diagrams of the homogeneous Bose-Hubbard model
in a 3D cubic lattice with total particle filling $n=1$ and $n=2$,
which feature superfluid and Mott-insulating phases and also notably
reveal ordered phases with XY-ferromagnetism and
N\'eel-antiferromagnetism in the Mott domain. We investigate the
critical temperatures of these long-range ordered states.
Moreover we have confirmed the stability of these magnetic phases in
a trapped 2D or 3D system. In the case of a 3D cubic lattice, we
have in particular computed XY-ferromagnetic ordering at finite
temperatures which is relevant for current experiments
\cite{Weld2009, Medley2010}. Another important issue is the
detection of novel magnetic phases with long-range spin order. For
an antiferromagnetic phase, spin-sensitive detection can be used to
detect the N\'eel-type order, i.e. one spin component can be
detected after removing the other one using spin-selective
single-site addressing, as pointed recently
\cite{Weitenberg2011,Kuhr2011}. For the XY-ferromagnetic phase, one
could apply a $\pi/2$ pulse with a position-dependent phase (linear
gradient), thus probing locally with different phases. For
XY-ferromagnetic long-range order one would thus observe stripes
with regular spacing when the phase matches and the $\pi/2$ pulse
transfers all atoms into one of the two spin states
\cite{Kuhr2011_1}.

Achieving the necessary ultra-low temperatures for detecting
magnetic ordering of cold bosons in optical lattices has so far
remained elusive. However, it is anticipated that by further
experimental advances this obstacle will be overcome in the near
future \cite{DeMarco2010}. Our results provide theoretical
benchmarks which in the future will also be extended to other
geometries such as triangular or hexagonal lattices.

\begin{acknowledgments}
We acknowledge useful discussions with S. Kuhr, I. Titvinidze, C.
Weitenberg and D. Weld. This work was supported by the China
Scholarship Fund (Y.L) and the Deutsche Forschungsgemeinschaft via
SFB-TR 49, FOR 801 and the DIP project BL 574/10-1.
\end{acknowledgments}

\end{document}